\documentstyle[12pt]{article}
\newtheorem{Lemma}{Lemma}
\newtheorem{Theorem}{Theorem}
\newtheorem{Corollary}[Theorem]{Corollary}
\def\proof{\par{\it Proof}. \ignorespaces}
\def\endproof{{\ \vbox{\hrule\hbox{%
   \vrule height1.3ex\hskip0.8ex\vrule}\hrule }}\par}
\newenvironment{Proof}{\proof}{\endproof}

\begin{document}

\title{Explicit Integration of the Full Symmetric Toda Hierarchy
and the Sorting Property}

\author{Y. KODAMA and K. T-R McLAUGHLIN
 \\{\it Department of Mathematics,
Ohio State University,}
\\ {\it Columbus, OH 43210, USA} \\
}

\date{February 22, 1995}

\maketitle

\begin{abstract}

We give an explicit formula for the solution to the initial value
problem of the full  symmetric Toda hierarchy.  The formula is obtained
by the orthogonalization procedure of Szeg\"{o}, and is also
interpreted as a consequence of the QR factorization method of Symes
\cite{symes}.  The sorting property of the dynamics is also proved
for the case of a generic symmetric matrix in the sense described in the text,
and generalizations of tridiagonal formulae are given for the
case of matrices with $2M+1$ nonzero diagonals.

\end{abstract}

\medskip


\par\medskip\medskip


\section{Introduction}
The finite non-periodic Toda lattice hierarchy can be
written in the Lax form \cite{flash}, \cite{moser}
\begin{eqnarray}
\label{toda}
\frac{\partial }{\partial t_{n}} L = \left[ B_{n} \ , \ L \right] \ ,
\ n = 1, 2, \cdots
\end{eqnarray}
where $L$ is a symmetric ``tridiagonal'' matrix, and $B_{n}$ is the
skew symmetric matrix obtained from $L$ by
\begin{eqnarray}
\label{skew}
B_{n} = \prod_{a} L^{n} \equiv \left( L^{n} \right)_{>0} -
\left( L^{n} \right)_{<0} \ ,
\end{eqnarray}
where $\left( L^{n} \right)_{>0 \ (<0)}$ denotes the strictly upper (lower)
triangular part of $L^{n}$.

An immediate consequence of (\ref{toda}) is that the eigenvalues of
$L$ are invariant with respect to the time ${\bf t} \equiv \left({t_{1}, t_{2},
\cdots} \right)$, and the Toda equation, corresponding to the case
$n=1$, as well as the hierarchy (\ref{toda}), can be shown to be a
completely integrable system.  Moser \cite{moser} showed that for the
Toda lattice $(n=1)$ the matrix $L$ converges as $t_{1}
\rightarrow \infty$ to a diagonal matrix containing the eigenvalues of
the matrix $L(t_{1}=0)$, arranged in decreasing order.  This sorting
property has
been further studied in
\cite{DNT}, where a general framework for calculating the eigenvalues
of a symmetric matrix using the Toda equations was developed.

In \cite{DNLT}, Deift et. al. extended the equation (\ref{toda}) to a
system where $L$ is an arbitrary symmetric matrix, which we call the
full symmetric Toda equation.  They then proved its complete
integrability by constructing the constants of motion and the
associated action - angle variables as a Hamiltonian system.

In this paper we give an explicit formula for the solution to the
initial value problem of the full symmetric Toda hierarchy, and show
that the sorting property also holds in this case.  The formulae we
construct for the eigenfunctions of $L$ turn out to be an
orthogonalization procedure described in Szeg{\"{o}}'s book
\cite{Sz39}.  This result is also obtained via the QR factorization
solution method of \cite{symes}.  We also construct alternative
formulae for the cases of symmetric $2M+1$ diagonal matrices, {\it i.e.}
 $a_{ij} =0$ for $|i-j| \ge M+1$.

\section{Full Symmetric Toda Hierarchy} The full symmetric Toda hierarchy
(\ref{toda}) can also
be written as the compatibility condition of the set of two linear
equations for the orthonormalized eigenvectors of the $N \times N$
symmetric matrix $L$,
\begin{eqnarray}
\label{comp1}
L \phi &=& \lambda \phi \ , \\
\label{comp2}
\frac{ \partial }{\partial t_{n}} \phi &=& B_{n} \phi \ .
\end{eqnarray}
Let us here give some properties of the eigenvalue problem
(\ref{comp1}).  Since the matrix $L$ is symmetric, it can be
diagonalized by an orthogonal matrix $\Phi \in O(N)$, i.e.
\begin{eqnarray}
\label{diag}
L = \Phi \ diag \left( \lambda_{1} , \ \cdots \ , \lambda_{N} \right)
\Phi^{T} \ .
\end{eqnarray}
Here the matrix $\Phi$ contains the eigenvectors of $L$,
say $\phi(\lambda_{k}) \equiv ( \phi_{1}(\lambda_{k}), \ \cdots \ $,
$\phi_{N}(\lambda_{k} ))^{T} $ for $k = 1, 2, \cdots , N$,
\begin{eqnarray}
\label{orth}
\Phi \equiv \left[ \phi(\lambda_{1}), \ \cdots \ , \ \phi(\lambda_{N})
\right] \ = \ \left[ \phi_{i}(\lambda_{j}) \right]_{1 \le i , j \le N}
 \ .
\end{eqnarray}
Since $\Phi \in O(N)$, we also have the orthogonality relations
\begin{eqnarray}
\label{ortho}
\sum_{k=1}^{N} \phi_{i}(\lambda_{k}) \phi_{j}(\lambda_{k}) =
\delta_{ij} \ , \\
\sum_{k=1}^{N} \phi_{k}(\lambda_{i}) \phi_{k}(\lambda_{j}) =
\delta_{ij} \ .
\end{eqnarray}
Then the entries of $L$, which we call $a_{ij}$, can be expressed as
\begin{eqnarray}
a_{ij} = < \lambda \phi_{i} \phi_{j} > \equiv \sum_{k=1}^{N}
\lambda_{k} \phi_{i}(\lambda_{k}) \phi_{j}(\lambda_{k}) \ .
\end{eqnarray}
Here the bracket $<>$ is defined as a linear functional on
${\bf F}$, the integrable functions of $\lambda$, with the Dirac
measure, i.e.
\begin{eqnarray}
\left.
\begin{array}{c}
< \ > \ : {\bf F} \rightarrow {\bf R} \\
\hspace{1.15in}f(\lambda) \mapsto  \int_{\bf R} f(\lambda) d \alpha(\lambda) \
, \\
\end{array}
\right\}
\end{eqnarray}
where the measure $d \alpha(\lambda)$ is
\begin{eqnarray}
d \alpha (\lambda) = \sum_{k=1}^{N} \delta(\lambda - \lambda_{k}) d
\lambda \ .
\end{eqnarray}
Note also that the entries of $L^{n}$, which we will call
$a_{ij}^{(n)}$, are given by
\begin{eqnarray}
a_{ij}^{(n)} \equiv \left( L^{n} \right)_{ij} = < \lambda^{n} \phi_{i}
\phi_{j} > \ .
\end{eqnarray}

To solve the equations (\ref{comp1}) and (\ref{comp2}), we first gauge
transform the eigenfunction $\phi(\lambda,{\bf t})$ by the diagonal matrix
\begin{eqnarray}
\nonumber
\Lambda = diag \big( < \psi_{1}^{2}>^{-1/2}\ , \cdots \ , \
< \psi_{N}^{2}>^{-1/2} \big) \ ,
\end{eqnarray}
\begin{eqnarray}
\label{trans}
\phi = \Lambda \psi \ .
\end{eqnarray}
Note that the transformation (\ref{trans}) includes a freedom in the
choice of $\psi$, i.e. (\ref{trans}) is invariant under $\psi_{i}
\rightarrow f_{i}({\bf t}) \psi_{i}$, with $\{f_{i}\}_{i=1}^{N}$ arbitrary
functions of ${\bf t}$.  With (\ref{trans}), (\ref{comp1}) and (\ref{comp2})
become
\begin{eqnarray}
\label{lam1}
\big(
\Lambda^{-1} L \Lambda \big) \psi &=& \lambda \psi \ , \\
\label{lam2}
\frac{ \partial }{\partial t_{n}} \psi &=&
\big( \Lambda^{-1} B_{n} \Lambda \big) \psi -
\left(\frac{ \partial }{\partial t_{n}} \log{\Lambda} \right) \psi \ .
\end{eqnarray}

Writing
\begin{eqnarray}
\nonumber
\left( \Lambda^{-1} B_{n} \Lambda \right) &=&
 -2 \left( \Lambda^{-1} L^{n} \Lambda \right)_{<0} +
\left( \Lambda^{-1} L^{n} \Lambda \right) - diag\left( L^{n} \right) \ ,
\end{eqnarray}
equation (\ref{lam2}) becomes
\begin{eqnarray}
\label{nlam}
\frac{ \partial }{\partial t_{n}} \psi &=&
 -2 \left( \Lambda^{-1} L^{n} \Lambda \right)_{<0} \psi +
\lambda^{n} \psi - \left( diag\left( L^{n} \right) +
\frac{ \partial }{\partial t_{n}} \log{ \Lambda} \right) \psi \ .
\end{eqnarray}
We here observe that (\ref{nlam}) can be split into two sets of
equations by fixing the freedom in the determination of $\psi$.  In
the components, these are
\begin{eqnarray}
\label{cpn1}
\frac{ \partial }{\partial t_{n}} \psi_{i} &=&
- 2 \sum_{j=1}^{i-1} \frac{ < \lambda^{n} \psi_{i} \psi_{j} >}
{< \psi_{j}^{2} >} \psi_{j} + \lambda^{n} \psi_{i} \ , \\
\label{cpn2}
\frac{1}{2} \frac{ \partial }{\partial t_{n}} \log{
< \psi_{i}^{2} > } &=& a_{ii}^{(n)} \ .
\end{eqnarray}
It is then easy to see that we have:
\begin{Lemma}
The solution of (\ref{cpn1}) can be written in the form of separation
of variables,
\begin{eqnarray}
\label{sepa}
\psi(\lambda;{\bf t}) = A({\bf t}) \phi^{0}(\lambda)
e^{\xi(\lambda,{\bf t})} \ ,
\end{eqnarray}
where $A({\bf t})$ is a lower triangular matrix with $diag(A({\bf t}))
= A({\bf t}=0) = I_{N}$, the $N \times N$ identity matrix,
$\phi^{0}({\bf t}) = \phi(\lambda;{\bf t})$, and $\xi(\lambda,{\bf t})
= \sum_{n=1}^{\infty} \lambda^{n} t_{n}$.
\end{Lemma}
Note that we have chosen the initial conditions of $\psi(\lambda,{\bf t})$
to coincide with $\phi(\lambda,{\bf t})$, $\psi(\lambda;0) \equiv
\phi^{0}(\lambda)$, and thus $<\psi_{i} \psi_{j}>({\bf t}=0) = <\phi_{i}^{0}
\phi_{j}^{0} > =
\delta_{ij}$.  As a direct consequence of this lemma, and the
orthogonality of the eigenvectors (\ref{ortho}), i.e. $< \psi_{i}
\psi_{j} > =0$ for $i \neq j$, we have:
\begin{Corollary}(Orthogonality):
For each $i \in \{2, \cdots, N\}$, we have for all ${\bf t}$ with
$t_{m} \in {\bf R}$,
\begin{eqnarray}
\label{corr}
<\psi_{i} \phi_{j}^{0} e^{\xi(\lambda,{\bf t})}> \equiv \sum_{k=1}^{N}
\psi_{i}(\lambda_{k};{\bf t}) \phi_{j}^{0}(\lambda_{k})
e^{\xi(\lambda_{k},{\bf t})} = 0
\end{eqnarray}
for $j = 1, 2, \cdots, i-1$.
\end{Corollary}

Now we obtain the main theorem of the section, which yields the formula
for the eigenvectors of $L$ in terms of the initial data $\{
\phi_{i}^{0}(\lambda) \}_{1 \le i
\le N}$:
\begin{Theorem}
\label{thth}
The quantities $\psi_{i}(\lambda; {\bf t})$ are given by
\begin{eqnarray}
\label{psis}
\psi_{i}(\lambda;{\bf t}) = \frac{e^{\xi(\lambda,{\bf t})}}
{ D_{i-1}({\bf t})} \left|
\begin{array}{ccc}
c_{11} & \ldots & c_{1i} \\
\vdots & \ddots & \vdots \\
c_{i-1 1} & \ldots & c_{i-1 i} \\
\phi_{1}^{0}(\lambda) & \ldots & \phi_{i}^{0}(\lambda) \\
\end{array}
\right|
\end{eqnarray}
where $c_{ij}({\bf t}) = < \phi_{i}^{0} \phi_{j}^{0} e^{ 2
\xi(\lambda,{\bf t})}>$, and $D_{k}({\bf t})$ is the determinant of the
$k \times k$ matrix with entries $c_{ij}({\bf t})$, i.e.
\begin{eqnarray}
\label{DDD}
D_{k}({\bf t}) = \left| \Big( c_{ij}({\bf t}) \Big)_{1 \le i,j \le k}
\right| \ .
\end{eqnarray}
 \end{Theorem}
\begin{Proof}
{}From equation (\ref{corr}) with (\ref{sepa}), we have
\begin{eqnarray}
\label{pro}
\sum_{ k = 1}^{i} A_{ik}({\bf t}) < \phi_{k}^{0} \phi_{\ell}^{0} e^{2
\xi(\lambda,{\bf t})} > = 0 \ , \ for \ 1 \le \ell \le i-1 \ .
\end{eqnarray}
Solving (\ref{pro}) for $A_{ik}$ with $A_{ii} = 1$, we obtain
\begin{eqnarray}
A_{ik}({\bf t}) = - \frac{ D_{i-1}^{k}({\bf t})}{D_{i-1}({\bf t})}
\end{eqnarray}
where $D_{i-1}^{k}({\bf t})$ is the determinant $D_{i-1}({\bf t})$ in the form
(\ref{DDD}) with the replacement of the kth column $\left( c_{1k}, \ldots \ ,
c_{i-1 \ k} \right)^{T}$ by $\left( c_{1i}, \ldots \ ,
c_{i-1 \ i} \right)^{T}$.  From (\ref{sepa}), we have
\begin{eqnarray}
\nonumber
\psi_{i} &=& e^{\xi(\lambda,{\bf t})} \sum_{k=1}^{i} A_{ik}\phi_{k}^{0} \\
\nonumber
&=& - e^{\xi(\lambda,{\bf t})}\sum_{k=1}^{i-1} \frac{D_{i-1}^{k}
\phi_{k}^{0}}{D_{i-1}} + e^{\xi(\lambda,{\bf t})}
\frac{D_{i-1} \phi_{i}^{0}}{D_{i-1}} \\
\nonumber
&=& \frac{e^{\xi(\lambda,{\bf t})}}{D_{i-1}} \times \left\{
 - \phi_{1}^{0}
\left|
\begin{array}{cccc}
c_{1i} & c_{21} & \cdots & c_{1 i-1} \\
\vdots & \ddots & \ddots & \vdots \\
c_{i-1 i} & c_{i-1 2} & \cdots & c_{i-1 i-1} \\
\end{array} \right|
+ \cdots \right.  \\
\nonumber
& &  \hspace{.3in} \left. \cdots +
 \phi_{i}^{0}
\left|
\begin{array}{ccc}
c_{11} & \cdots & c_{1 i-1} \\
\vdots & \ddots & \vdots \\
c_{1 i-1} & \cdots & c_{i-1 i-1} \\
\end{array} \right| \right\} \ ,
\end{eqnarray}
which one can see is just (\ref{psis}).
\end{Proof}

We note from (\ref{psis}) that $<\psi_{i}^{2}>$ can be expressed by
$D_{i}$,
\begin{eqnarray}
< \psi_{i}^{2} > = \frac{D_{i}}{D_{i-1}} \ .
\end{eqnarray}
This yields the formulae for the normalized eigenfunctions
\begin{eqnarray}
\label{evcs}
\phi_{i}(\lambda,{\bf t}) = \frac{e^{\xi(\lambda,{\bf t})}}
{\sqrt{D_{i}({\bf t})
D_{i-1}({\bf t})}} \left|
\begin{array}{ccc}
c_{11} & \ldots & c_{1i} \\
\vdots & \ddots & \vdots \\
c_{i-1 1} & \ldots & c_{i-1 i} \\
\phi_{1}^{0}(\lambda) & \ldots & \phi_{i}^{0}(\lambda) \\
\end{array}
\right|
\end{eqnarray}

This derivation of (\ref{evcs}) is nothing but the orthogonalization
procedure of Szeg{\"{o}}, which is equivalent to Gram - Schmidt
orthogonalization.  Indeed, using the fact that $\phi_{i} =
\frac{\psi_{i}} {< \psi_{i}^{2} >^{1/2}}$, we see that equation
(\ref{sepa}) expresses the relations
\begin{eqnarray}
\label{spa}
\phi_{i}(\lambda,{\bf t}) \in span \left\{
\phi_{1}^{0}(\lambda)e^{\xi(\lambda,{\bf t})}, \cdots,
\phi_{i}^{0}(\lambda)e^{\xi(\lambda,{\bf t})} \right\} \ , \ i=1, \cdots, N,
\end{eqnarray}
where we are viewing $\{ \phi_{i}(\lambda,{\bf t}) \}_{i=1}^{N}$ as a
collection of orthogonal functions defined on the eigenvalues of the
matrix $L$.  Relations (\ref{spa}), together with the orthogonality
\begin{eqnarray}
\label{ort2}
<\phi_{i} \phi_{j}> = \delta_{ij} \ ,
\end{eqnarray}
imply that $\left\{ \phi_{i}(\lambda,{\bf t})
\right\}_{i=1}^{N}$ are obtained by orthogonalization of the sequence
$\left\{ \phi_{i}^{0}(\lambda)e^{\xi(\lambda,{\bf t})} \right\}_{i=1}^{N}$,
and hence one may obtain equation (\ref{evcs}) from classical
formulae, directly from (\ref{spa}) and (\ref{ort2}).

With formula (\ref{evcs}), one can solve the initial value problem of
(\ref{comp1}) and (\ref{comp2}), via
\begin{eqnarray}
a_{ij}({\bf t}) = < \lambda \phi_{i} \phi_{j} >({\bf t}) \ .
\end{eqnarray}

We also remark that the determinant $D_k$ in (12) gives the $tau$-function of
this hierarchy, and plays a fundamental role in integrable systems
\cite{naka}.

\section{The QR Factorization Method} Here we give an alternative derivation of
the formula
(\ref{evcs}) by using the QR factorization method of Symes \cite{symes}.

Let $L^{0}$ be the initial matrix of $L({\bf t})$, i.e. $L(0) = L^{0}$.
Symes' factorization method is to compute the unique decomposition of
the symmetric matrix
\begin{eqnarray}
\label{syo}
e^{\xi(L^{0};{\bf t})} = Q^{T}({\bf t})R({\bf t})
\end{eqnarray}
where $\xi (L^{0};{\bf t}) = \sum_{n=1}^{\infty}t_{n} \left( L^{0}
\right)^{n}$, $Q({\bf t}) \in O(N)$ with $Q(0) = I_{N}$ and $R({\bf t})$
is an upper triangular matrix with $R_{ii} >0$  Then the evolution of
$L({\bf t})$ can be obtained by the relation
\begin{eqnarray}
\label{syt}
L({\bf t}) = Q ({\bf t}) L^{0} Q^{T}({\bf t}).
\end{eqnarray}
The orthogonal matrix $\Phi({\bf t})$ defined in (\ref{orth}) is then given
by
\begin{eqnarray}
\Phi({\bf t}) = Q({\bf t}) \Phi^{0}
\end{eqnarray}
where $\Phi^{0} = \big( \phi_{i}^{0}(\lambda_{j}) \big)_{1 \le i,j
\le N}$.

Now from the diagonalization of $L^{0}$ in (\ref{diag}), we have
\begin{eqnarray}
e^{\xi(L^{0};{\bf t})} = \Phi^{0} diag \left( e^{\xi(\lambda_{1};{\bf t})},
\cdots, e^{\xi(\lambda_{N};{\bf t})} \right) \left( \Phi^{0} \right)^{T}
\end{eqnarray}
which, together with (\ref{syo}) and (\ref{syt}) leads to
\begin{eqnarray}
\label{szz}
diag \left( e^{\xi(\lambda_{1};{\bf t})},
\cdots, e^{\xi(\lambda_{N};{\bf t})} \right)
\left( \Phi^{0} \right)^{T} R({\bf t})^{-1}
= \Phi^{T}({\bf t})  \ .
\end{eqnarray}
Equation (\ref{szz}) should be compared with (\ref{sepa}); we note
that (\ref{szz}) immediately implies relations (\ref{spa}).  Hence, as
described following relations (\ref{spa}), the factorization method of
Symes \cite{symes} is enough to establish Theorem \ref{thth}, and in
particular formula (\ref{evcs}).

\section{Sorting Property}
Here we consider the case of the full symmetric Toda equation, that
is, we have only one time $t_{1}$, which we denote by $t$.
As a consequence of the explicit formula for $\phi_{i}(\lambda;{\bf t})$ in
Theorem \ref{thth}, we can now show the sorting property, i.e. as $t
\rightarrow \infty$, the matrix $L(t)$ converges to a diagonal matrix
with ordered eigenvalues.  In order to show this, we first show:
\begin{Lemma}
\label{DAS}
Let the eigenvalues of $L$ be ordered as $\lambda_{1} > \lambda_{2} >
\cdots > \lambda_{N}$.  Then, as $t \rightarrow \infty$, we have
\begin{eqnarray}
D_{n}(t) \exp{\left( - 2 \sum_{i = 1}^{N} \lambda_{i} t \right)}
\rightarrow \left( det \Phi_{n}^{0} \right)^{2}
\end{eqnarray}
where $\Phi_{n}^{0}$ is the $n \times n$ matrix given by
$\Phi_{n}^{0} = \left[ \left( \phi_{i}^{0}(\lambda_{j}) \right)_{1 \le
i,j \le n} \right]$.
\end{Lemma}
\begin{Proof}
{}From (\ref{DDD}), we have
\begin{eqnarray}
\label{DDS}
D_{n}=\sum_{\lambda_{1} = 1}^{N} \cdots \sum_{\lambda_{n}=1}^{N}
\prod_{i=1}^{n}\phi_{i}^{0}(\lambda_{i}) e^{2 t \lambda_{i}}
\left|
\begin{array}{ccc}
\phi_{1}^{0}(\lambda_{1}) & \cdots & \phi_{1}^{0}(\lambda_{n}) \\
\vdots & \ddots & \vdots \\
\phi_{n}^{0}(\lambda_{1}) & \cdots & \phi_{n}^{0}(\lambda_{n}) \\
\end{array} \right| \ .
\end{eqnarray}
Expanding this in terms of $\exp{ \{ 2 ({\bf \ell},
\lambda) t \} }$ with $({\bf \ell}, \lambda) = \sum_{i=1}^{N} \ell_{i}
\lambda_{i}$ for $\ell_{i} \in {\bf Z}_{\ge 0}$, and with $\left| {\bf \ell}
\right| \equiv \sum_{i=1}^{N} \ell_{i} = n$, we obtain, setting $x_{i}
\equiv \exp{ \left( 2 \lambda_{i} t \right)}$,
\begin{eqnarray}
\label{DN}
D_{n}(t) = \sum_{\left| {\bf \ell} \right| = n} d({\bf \ell}) x_{1}^{\ell_{1}}
\cdots x_{N}^{\ell_{N}} \ .
\end{eqnarray}
We first show that the nonvanishing dominant term (for $t$ large)
in (\ref{DN}) is given by the case ${\bf \ell} = ( 1, \ldots, 1, 0, \ldots,
0)$, that is, $\ell_{i} = 1$, for $ i = 1, \ldots, n$, and $\ell_{i} =
0 $ otherwise.

To show this, we note that
\begin{eqnarray}
d({\bf \ell}) = 0 \
\end{eqnarray}
if there exists $\ell_{i} \ge 2$ for some $i$.  Indeed, for such a
term, the determinant in (\ref{DDS}) would necessarily vanish.  This,
along with the ordering of the eigenvalues,
implies that the dominant term in (\ref{DN}) is the one with ${\bf \ell} = (
1, \ldots, 1, 0, \ldots, 0)$, and is given by
\begin{eqnarray}
d( 1, \ldots, 1, 0, \ldots,0) = \frac{ \partial^{n}}{\partial x_{1}
\cdots \partial  x_{n}} D_{n}(t) \Big|_{x_{1} = \cdots = x_{N} = 0} \ .
\end{eqnarray}
Which is, summing over all permutations
\begin{eqnarray}
\nonumber
{\bf P}_{n} =
\left(
\begin{array}{cccc} 1& 2& \cdots &n \\
\ell_{1}& \ell_{2}& \cdots &\ell_{n} \\
\end{array} \right) \ ,
\end{eqnarray}
\begin{eqnarray}
\nonumber
d( 1, \ldots, 1, 0, \ldots, 0) &=& \sum_{{\bf P}_{n}}
\left|
\begin{array}{ccc}
\phi_{1}^{0}(\lambda_{\ell_{1}}) \phi_{1}^{0}(\lambda_{\ell_{1}}) , &
\cdots & \phi_{1}^{0}(\lambda_{\ell_{1}})
\phi_{n}^{0}(\lambda_{\ell_{1}}) \\
\vdots & \ddots & \vdots \\
\phi_{1}^{0}(\lambda_{\ell_{n}}) \phi_{n}^{0}(\lambda_{\ell_{n}}) , &
\cdots & \phi_{n}^{0}(\lambda_{\ell_{n}})
\phi_{n}^{0}(\lambda_{\ell_{n}}) \\
\end{array}
\right| \\
\nonumber
&=&
\sum_{{\bf P}_{n}} \phi_{1}^{0}(\lambda_{\ell_{1}}) \cdots
\phi_{n}^{0}(\lambda_{\ell_{n}})
\left|
\begin{array}{ccc}
\phi_{1}^{0}(\lambda_{\ell_{1}})  &
\cdots & \phi_{n}^{0}(\lambda_{\ell_{1}}) \\
\vdots & \ddots & \vdots \\
\phi_{1}^{0}(\lambda_{\ell_{n}})  &
\cdots & \phi_{n}^{0}(\lambda_{\ell_{n}}) \\
\end{array}  \right| \\
&=& \left( det \Phi_{n}^{0} \right)^{2} \ .
\end{eqnarray}
\end{Proof}

Now we have
\begin{Theorem}
Suppose that $det \Phi_{n}^{0} \neq 0$ for $ n = 1, \ldots , N$.  Then
as $t \rightarrow \infty$, the eigenfunctions
$\phi_{i}(\lambda_{j};t)$ satisfy
\begin{eqnarray}
\label{dell}
\phi_{i}(\lambda_{j}) \rightarrow \delta_{ij} \times
sgn \left( det \Phi_{i}^{0}\right) sgn\left( det \Phi_{i-1}^{0} \right)  \ ,
\end{eqnarray}
which implies $L(t) \rightarrow diag \left(\lambda_{1}, \ldots ,
\lambda_{N} \right)$, from (\ref{diag}).
\end{Theorem}
\begin{Proof}
{}From the formula (\ref{evcs}) and Lemma \ref{DAS} we have, as $t
\rightarrow \infty$,
\begin{eqnarray}
\nonumber
& &\phi_{n}(\lambda; t) \rightarrow \frac{ e^{\lambda t}}
{\left| det \Phi_{n}^{0} \right| \left| det \Phi_{n-1}^{0} \right| \exp{
\left[ \left( 2 \sum_{i=1}^{n-1} \lambda_{i} + \lambda_{n} \right) t
\right]}} \times \\
\label{blt}
& & \hspace{.6in} \times  \left|
\begin{array}{ccc}
c_{11}(t) & \cdots & c_{1n}(t) \\
\vdots & \ddots & \vdots \\
c_{1 n-1}(t) & \cdots & c_{n-1 n} \\
\phi_{1}^{0}(\lambda) & \cdots & \phi_{n}^{0}(\lambda) \\
\end{array} \right| \ .
\end{eqnarray}
The dominant term in the determinant gives
\begin{eqnarray}
\nonumber
& &
e^{2 \sum_{i=1}^{n-1} \lambda_{i}t} \sum_{{\bf P}_{n-1}}
\phi_{1}^{0}(\lambda_{\ell_{1}}) \cdots
\phi_{n-1}^{0}(\lambda_{\ell_{n-1}})
\left|
\begin{array}{ccc}
\phi_{1}^{0}(\lambda_{\ell_{1}}) & \ldots &
\phi_{n}^{0}(\lambda_{\ell_{1}}) \\
\vdots & \ddots & \vdots \\
\phi_{1}^{0}(\lambda_{\ell_{n-1}}) & \ldots &
\phi_{n}^{0}(\lambda_{\ell_{n-1}}) \\
\phi_{1}^{0}(\lambda) & \ldots &
\phi_{n}^{0}(\lambda) \\
\end{array} \right| \\
\label{thdo}
& & =
e^{2 \sum_{i=1}^{n-1} \lambda_{i}t}
\left|
\begin{array}{ccc}
\phi_{1}^{0}(\lambda_{1}) & \ldots &
\phi_{n}^{0}(\lambda_{1}) \\
\vdots & \ddots & \vdots \\
\phi_{1}^{0}(\lambda_{n-1}) & \ldots &
\phi_{n}^{0}(\lambda_{n-1}) \\
\phi_{1}^{0}(\lambda) & \ldots &
\phi_{n}^{0}(\lambda) \\
\end{array} \right| \times \\
\nonumber
& & \hspace{1.0in} \times
\sum_{{\bf P}_{n-1}}
\phi_{1}^{0}(\lambda_{\ell_{1}}) \cdots
\phi_{n-1}^{0}(\lambda_{\ell_{n-1}}) \#(\{ \ell_{j} \}_{j=1}^{n-1}) \ ,
\end{eqnarray}
where $\#(\{ \ell_{j} \}_{j=1}^{n-1})$ is the signature of the
permutation $\{ \ell_{j} \}_{j=1}^{n-1}$.  We note that the
determinant in (\ref{thdo}) is
zero for $\lambda = \lambda_{j}$, $j=1, \ldots, n-1$.  The ordering
$\lambda_{1} > \cdots > \lambda_{N}$ implies the result.
\end{Proof}

Observe that the arbitrariness in the selection of eigenvectors of the
initial matrix appears in the asymptotic formula (\ref{dell}).  However,
 independent of this choice (naturally), we see from Theorem 3
that for any generic symmetric initial matrix, that is, $L_{0}$ satisfies
$\Phi_{j}^{0} \neq 0$ for $j=1, \ldots, N$, and has distinct
eigenvalues, the Toda lattice converges to a diagonal matrix,
containing the eigenvalues arranged in decreasing order.

\section{Finite-Band Toda Hierarchy} As a final comment, we give alternative
formulae for
$\phi_{i}(\lambda,{\bf t})$, for the case where the matrix $L$ has $(2M +1)$
nonzero diagonals, $M=1$ (tridiagonal), $M=2$ (pentadiagonal), and so
on.  Thus we have $a_{ij}=0$ for $|i-j|>M$.  We further assume $a_{ij} \neq 0$
 for $|i-j|=M$, and that the eigenvalues of $L$ are distinct.

For such a $2M+1$ - diagonal matrix, the first column of the
eigenvector equation $L \phi = \lambda \phi$ is
\begin{eqnarray}
\label{eve}
\sum_{i=1}^{M}a_{1i} \phi_{i}(\lambda) + a_{1 \
M+1}\phi_{M+1}(\lambda) \ = \ \lambda \phi_{1}(\lambda) \ .
\end{eqnarray}
When the eigenvectors $\phi(\lambda)$ are viewed as functions over the
eigenvalues $\{ \lambda_{j}\}_{j=1}^{N}$, equation (\ref{eve})
expresses the relation
\begin{eqnarray}
\phi_{M+1}(\lambda) \in span \left\{ \phi_{1}(\lambda), \ldots,
\phi_{M}(\lambda), \lambda \phi_{1}(\lambda) \right\} \ .
\end{eqnarray}
Likewise, we have
\begin{eqnarray}
\phi_{M+j}(\lambda) \in span \left\{ \phi_{1}(\lambda), \ldots,
\phi_{M}(\lambda), \lambda \phi_{1}(\lambda), \ldots ,
\lambda \phi_{j}(\lambda) \right\}
\end{eqnarray}
for $1 \le j \le M$.  We leave the reader to verify that we have
\begin{eqnarray}
\label{dspa}
& & \phi_{kM + j}(\lambda) \in span \left\{\phi_{1}(\lambda), \ldots,
\phi_{M}(\lambda), \lambda \times \left[ \phi_{1}(\lambda), \ldots,
\phi_{M}(\lambda)\right], \ldots \right. \hspace{.4in} \\
\nonumber
& & \hspace{2.0in} \left.
\ldots,  \lambda^{k} \times
\left[ \phi_{1}(\lambda), \ldots, \phi_{j}(\lambda)\right] \right\} \ , \\
\nonumber
& & \hspace{3.in}
\begin{array}{c}
k = 0, 1, \ldots \\
j = 1, \ldots, M \\
\end{array} \ .
\end{eqnarray}
Relations (\ref{dspa}), together with the orthogonality (\ref{ort2}),
yield the following orthogonalization procedure.

Define, for $k = 0, 1, \ldots \ $and $j = 1, \ldots, M$,
\begin{eqnarray}
\rho_{kM + j} = \lambda^{k} \phi_{j}(\lambda) \ ,
\end{eqnarray}
and
\begin{eqnarray}
\hat{c}_{ij} = < \rho_{i} \rho_{j} > \ .
\end{eqnarray}
Then we have
\begin{eqnarray}
\phi_{i}(\lambda) = \frac{1}{\sqrt{D_{i-1}D_{i}}}
\left|
\begin{array}{ccc}
\hat{c}_{11} & \cdots & \hat{c}_{1i} \\
\vdots & \ddots & \vdots \\
\hat{c}_{i-1 \ 1} & \ldots & \hat{c}_{i-1 \ i} \\
\rho_{1} & \ldots & \rho_{i} \\
\end{array} \right| \ ,
\end{eqnarray}
where
\begin{eqnarray}
D_{n} = det \left[ \hat{c}_{ij} \right]_{1 \le i,j \le n} \ .
\end{eqnarray}

We remark at this point that if $a_{k_{0}M + i_{0} , \ (k_{0}+1)M +
i_{0}} = 0$ for some $1 \le i_{0} \le M-1$, $k_{0} \ge 1$, then the
procedure is modified by the replacement of $\rho_{kM+i_{0}}$ by
$\lambda^{k-k_{0}} \phi_{k_{0}M + i_{0}}$ for $k \ge k_{0}$.

Now from the preceding sections, when the matrix $L({\bf t})$ evolves
according to the Toda hierarchy, we have the relations
\begin{eqnarray}
\label{nsp}
\phi_{kM+j}(\lambda,{\bf t}) \in span \left\{
\phi_{1}^{0}(\lambda)e^{\xi(\lambda,{\bf t})}, \ldots,
\phi_{kM+j}^{0}e^{\xi(\lambda,{\bf t})} \right\} \ ,
\end{eqnarray}
for $k \ge 0$ and $1 \le j \le M$, which, together with
(\ref{dspa}) imply
\begin{eqnarray}
\phi_{kM+j}(\lambda,{\bf t}) \in span \left\{
\lambda^{0} \times \left[
\phi_{1}^{0}(\lambda)e^{\xi(\lambda,{\bf t})}, \ldots,
\phi_{M}^{0}(\lambda)e^{\xi(\lambda,{\bf t})} \right], \ldots \hspace{.3in}
 \right. \\
\nonumber
\hspace{2.in}
\left. \ldots,
\lambda^{k} \times \left[
\phi_{1}^{0}(\lambda)e^{\xi(\lambda,{\bf t})}, \ldots,
\phi_{j}^{0}(\lambda)e^{\xi(\lambda,{\bf t})} \right] \right\}
\end{eqnarray}
and hence for all ${\bf t}$ with $t_{i} >0$, we have the following
orthogonalization procedure.
Define
\begin{eqnarray}
\rho_{kM+j}(\lambda,{\bf t}) = \lambda^{k} \phi_{j}^{0} e^{\xi(\lambda,{\bf
t})}
 \ , \
\begin{array}{c}
k = 0, \ldots \\
j = 1, \ldots, M \\
\end{array} \ ,
\end{eqnarray}
and
\begin{eqnarray}
c_{ij}({\bf t}) = < \rho_{i} \rho_{j} >({\bf t}) \ .
\end{eqnarray}
Then we have
\begin{eqnarray}
\phi_{i}(\lambda,{\bf t}) = \frac{1}{\sqrt{D_{i-1}D_{i}}}
\left|
\begin{array}{ccc}
c_{11}({\bf t}) & \cdots & c_{1i}({\bf t}) \\
\vdots & \ddots & \vdots \\
c_{i-1 \ 1}({\bf t}) & \ldots & c_{i-1 \ i}({\bf t}) \\
\rho_{1}(\lambda,{\bf t}) & \ldots & \rho_{i}(\lambda,{\bf t}) \\
\end{array} \right| \ ,
\end{eqnarray}
where
\begin{eqnarray}
D_{n} = det \left[ c_{ij}({\bf t}) \right]_{1 \le i,j \le n} \ .
\end{eqnarray}

\par\medskip\medskip
{\bf Acknowledgment}
One of the authors (Y.K.) wishes to thank Prof. Y. Nakamura for
introducing him to a fascinating world of the finite Toda molecule.
The work of Y.K. is partially supported by an NSF grant DMS9403597.

A tridiagonal version of some of these arguments were shown to K. T-R
M.  by Prof. Percy Deift, in a course on mechanics at NYU, in 1990.

K. T-R M. wishes to thank Prof. Y. Kodama for guidance and for greatly
encouraging this collaboration, and to thank Prof. Percy Deift for helpful
guidance.


\bibliographystyle{amsplain}

\begin{thebibliography}{7}

\bibitem{DNLT}
P. Deift, L.C. Li, T. Nanda, and C. Tomei.
The Toda Flow on a Generic Orbit is Integrable. {\it Comm. Pure Apple. Math.},
{\bf 39} (1986), 183-232.

\bibitem{DNT}
P.Deift, T. Nanda, and C. Tomei.
Ordinary Differential Equations and the Symmetric Eigenvalue Problem.
{\it SIAM J. Numer. Anal.}, {\bf 20} (1983), 1-20.

\bibitem{flash}
H. Flaschka.  On the Toda Lattice II. {\it Prog. Theor. Phys.},
{\bf 51} (1974), 703-716.

\bibitem{moser}
J. Moser.   Finitely Many Mass Points on the Line Under the Influence
of an Exponential Potential - An Integrable System. in: Dynamical
Systems, Theory and Applications, J. Moser ed. {\it Lec. Notes in Phys.},
Vol.38, Springer-Verlag, Belrin and New York, 1975, pp. 467-497.

\bibitem{naka}
Y. Nakamura.  A tau-function for the finite Toda molecule,
and information spaces.  in: Symplectic Geometry and Quantizations, Y. Maeda,
H. Omori, and A. Weinstein eds., {\it Contemp. Math.} Amer. Math. Soc.
Providence, 1995,  to appear.

\bibitem{symes}
W.W. Symes.  The QR Algorithm and Scattering for the Finite Non-periodic
Toda Lattice. {\it Physica D}, {\bf 4} (1982), 275-278.

\bibitem{Sz39}
G. Szeg\"o. {\it Orthogonal Polynomials.} AMS Colloquium Publications,
vol.23, 1939.

\end{thebibliography}

\end{document}